# COMPUTER MODELLING OF RADIATIONALLY AND THERMALLY STIMULATED PROCESSES IN A CRYSTAL

K. Baktybekov, Ye. Vertyagina

*L.N. Gumilyov Eurasian National University, Astana, Kazakhstan*

The processes of radiation defects formation and evolution have been simulated in cubic dielectric crystals by the computational method of cellular automata. The ion crystal model is realized as two- and three-dimensional version by spatial size of $10^6$ lattice points. Irradiation duration is defined by the number of "generation-recombination" cycles and it has reached 2 million iteration. In the beginning of every cycle the concentration of radiation-induced defects in the lattice corresponds to the dose rate of irradiation.

As a result of performed numerical experiment it is obtained that the growth rate of primary defects concentration in a lattice is determined by the irradiation dose rate. If suppose that the defects concentration as a parameter, which characterizes a system state, reaches some critical value at the irradiation process, in this case there would be collective effects in the system that lead to self-organization phenomena and demonstrate a formation of one-type defect clusters. Concentration of radiation defects in clusters is higher than their average concentration over a crystal, so the creation of aggregate centers in the range of an accumulation of the same type radiation defects is more probably than in the event of homogeneous distribution of defects in a crystal bulk.

The fractal theory is used to detect synergetic effects and describe the spatial and temporal transformation of defect structure in a crystal. Calculation of fractal dimensions, function of multifractal spectrum and value of information entropy at the every stage of the modelling allows determining the fact that chaotic distribution of electron and hole centers is formed in the lattice at the short irradiation time, but multifractal one is created at the long irradiation.

To find out a mechanism of aggregate formation the simulation of the process of radiation defects accumulation have been performed in the square lattice by the size of 500x500 nodes at the 0,1% dose rate of irradiation for the following four cases.

1 – «black sphere 1», the probability of defects pair recombination $w = 1$, defects from the first coordination environment take part in the interaction process;

2 – «black sphere 2», defects from the first and the second coordination environment participate in interaction with equally probability, herewith the recombination probability is $w = 1$;

3 – «gray sphere 1», in this case the pairs recombination is realized with the probability $w = 0,75$, defects from the first coordination environment take part in the interaction process;

4 – «gray sphere 2», defects from the first and the second coordination environment participate in recombination process, recombination probability is equal to $w = 0,5$ at the interaction with the first coordination environment, and it is $w = 0,25$ with the second coordination environment.

Cases 1 and 3 in this simulation problem correspond to the physical process of interaction with short-distance recombination of radiation defects. Modelling the electron and hole centers interaction in the cases 2 and 4 it takes into account the probability of implementation of defects recombination by tunneling mechanism besides short-distance recombination.

On the base of calculation results the conclusion can be done that the accounting the interaction of electron and hole centers from the second coordination environment leads to the fact that the process of accumulation of radiation defects in a crystal is not observed. Aggregation only occurs in the case when a defect interacts with a center from the nearest surrounding and it is not irrespective of "gray" degree recombination environment. For the cases of "gray sphere" the higher rate of defects accumulation is intrinsic as opposed to "black sphere". It seems to connect with the fact that the single pair of different type defects, which are located in the neighbouring anion sites, has nonzero probability for survival at the time point $t$ under the condition of "gray sphere". Therefore, at the time point $t + \Delta t$ the number of particles in the system can be large in comparison with a system, which has a "black sphere" rule. In the event when the interaction of radiation defects from the second coordination environment is considering, i.e. there is the mechanism of tunneling recombination of electron and hole centers, apparently, the weakly nonequilibrium system is arisen, in which the stationary concentration of spatially homogeneous distributed defects is supported. In this case in consideration of calculated multifractal parameters it is evidently that a self-organization effect is absent in the system.